\documentclass[aip,
 mph,%
 amsmath,
 amssymb,
 superscriptaddress,
 floatfix,
 preprint,
 eqsecnum,%
 numerical,%
]{revtex4-1}

\usepackage{graphicx}
\DeclareGraphicsExtensions{.eps,.png,.jpg}
\usepackage[caption=false%,labelformat=parens,subrefformat=subparens,listofformat=subparens
]{subfig}
\usepackage{epsfig}
\usepackage{epstopdf}
\usepackage{lipsum}
\usepackage{wrapfig}
\usepackage{wasysym}
\usepackage{threeparttable,booktabs,siunitx}
\usepackage{booktabs}
\usepackage{lmodern}
\usepackage{longtable}
\usepackage{multirow}
\usepackage{amsmath}
\usepackage[OT2,T1]{fontenc}
\usepackage[bookmarks=false]{hyperref}
\usepackage{color}
\hypersetup{pdfauthor={},pdfborder=0 0 0,colorlinks=true,citecolor=blue,linkcolor=blue,urlcolor=blue,}

\DeclareSymbolFont{cyrletters}{OT2}{wncyr}{m}{n}
\DeclareMathSymbol{\comb}{\mathalpha}{cyrletters}{"58}

\newcommand{\ben}{\begin{eqnarray}\displaystyle}
\newcommand{\een}{\end{eqnarray}}

\begin{document}

\title{DIR-DBTnet: Deep iterative reconstruction network for 3D digital breast tomosynthesis imaging}
% Force line breaks with \\

\author{Ting Su}
 \thanks{Ting Su and Xiaolei Deng have made equal contributions to this work and both are considered as the first authors.}
\affiliation{Research Center for Medical Artificial Intelligence, Shenzhen Institutes of Advanced Technology, Chinese Academy of Sciences, Shenzhen, Guangdong 518055, China.}%Lines break automatically or can be forced with \\
\author{Xiaolei Deng}
 \thanks{Ting Su and Xiaolei Deng have made equal contributions to this work and both are considered as the first authors.}
\affiliation{Research Center for Medical Artificial Intelligence, Shenzhen Institutes of Advanced Technology, Chinese Academy of Sciences, Shenzhen, Guangdong 518055, China.}%Lines break automatically or can be forced with \\
\author{Zhenwei Wang}
\affiliation{Shanghai United Imaging Healthcare Co., Ltd, Shanghai 201807, China.}%
\author{Jiecheng Yang}%
 \affiliation{Research Center for Medical Artificial Intelligence, Shenzhen Institutes of Advanced Technology, Chinese Academy of Sciences, Shenzhen, Guangdong 518055, China.}
\author{Jianwei Chen}
\affiliation{Research Center for Medical Artificial Intelligence, Shenzhen Institutes of Advanced Technology, Chinese Academy of Sciences, Shenzhen, Guangdong 518055, China.}
\author{Hairong Zheng}%
\affiliation{Paul C Lauterbur Research Center for Biomedical Imaging, Shenzhen Institutes of Advanced Technology, Chinese Academy of Sciences, Shenzhen, Guangdong 518055, China.}%
\author{Dong Liang}%
 \thanks{Scientific correspondence should be addressed to Dong Liang (dong.liang@siat.ac.cn) and Yongshuai Ge (ys.ge@siat.ac.cn).}
 \affiliation{Research Center for Medical Artificial Intelligence, Shenzhen Institutes of Advanced Technology, Chinese Academy of Sciences, Shenzhen, Guangdong 518055, China.}
\affiliation{Paul C Lauterbur Research Center for Biomedical Imaging, Shenzhen Institutes of Advanced Technology, Chinese Academy of Sciences, Shenzhen, Guangdong 518055, China.}%
\author{Yongshuai Ge}%
 \thanks{Scientific correspondence should be addressed to Dong Liang (dong.liang@siat.ac.cn) and Yongshuai Ge (ys.ge@siat.ac.cn).}
 \affiliation{Research Center for Medical Artificial Intelligence, Shenzhen Institutes of Advanced Technology, Chinese Academy of Sciences, Shenzhen, Guangdong 518055, China.}
\affiliation{Paul C Lauterbur Research Center for Biomedical Imaging, Shenzhen Institutes of Advanced Technology, Chinese Academy of Sciences, Shenzhen, Guangdong 518055, China.}%

\date{\today}% It is always \today, today,

\begin{abstract}
\noindent {\bf Purpose:} The goal of this study is to develop a novel deep learning (DL) based reconstruction framework to improve the digital breast tomosynthesis (DBT) imaging performance.\\
%Digital breast tomosynthesis (DBT) is a promising imaging technique for providing 3D information of the breast. Due to the super-short tube scanning range, however, the quality of the reconstructed DBT images is usually degraded. 
{\bf Methods:} In this work, the DIR-DBTnet is developed for DBT image reconstruction by unrolling the standard iterative reconstruction algorithm within the deep learning framework. In particular, such network learns the regularizer and the iteration parameters automatically through network training with a large amount of simulated DBT data. Afterwards, both numerical and experimental data are used to evaluate its performance. Quantitative metrics such as the artifact spread function (ASF), breast density, and the signal difference to noise ratio (SDNR) are used for image quality assessment.  \\
{\bf Results:} For both numerical and experimental data, the proposed DIR-DBTnet generates reduced in-plane shadow artifacts and out-of-plane artifacts compared with the filtered back projection (FBP) and total variation (TV) methods. Quantitatively, the full width half maximum (FWHM) of the measured ASF curve from the numerical data is 33.4\% and 19.7\% smaller than those obtained with the FBP and TV methods, respectively; the breast density of the network reconstructed DBT images is more accurate and consistent with the ground truth. \\
{\bf Conclusions:} In conclusion, a deep iterative reconstruction network, DIR-DBTnet, has been proposed. Both qualitative and quantitative analyses of the numerical and experimental results show superior DBT imaging performance than the FBP and iterative algorithms. \\

\end{abstract}

\maketitle
             
\section{Introduction}
Recently, the X-ray digital breast tomosynthesis (DBT) technique becomes an important three-dimensional (3D) tomographic imaging method in detection and diagnosis of breast cancers. By collecting a sequence of two-dimensional (2D) projections within a limited scanning angular range, 3D breast images can be reconstructed. With the additional depth information, DBT is able to partially solve the tissue-overlapping problem encountered in digital mammography (DM), and thus enhancing the detection of abnormalities located in different planes \cite{TPlocaltumor2019}. However, the reconstructed DBT images may still suffer from the in-plane and out-of-plane artifacts due to the incompleteness of the acquired projection data. As a result, the lesion detection performance would be limited\cite{chen2012anatomical, chen2013association, doi:10.1002/mp.12864}.

To improve the DBT image quality, so far various reconstruction strategies have been proposed. For example, proper modifications to the standard ramp filter used in the filtered back projection (FBP) algorithm are able to generate improved mass detectability \cite{sechopoulos2013review, lu2015multiscale, rose2019filtered}. Additionally, the iterative reconstruction (IR) algorithms can also be used to deal with the ill-posed limited angle problem \cite{xu2015statistical, wu2004comparison, rose2017investigating} in DBT imaging. To do so, a certain objective function containing the data fidelity term and the regularization term is minimized. Regularizers such as total variation (TV) \cite{kastanis20083d, mota2015total}, total $p$ variation (TpV) \cite{sidky2009enhanced}, selective-diffusion regularization \cite{lu2010selective}, curvelet sparse regularization \cite{frikel2013sparse}, etc. \cite{zheng2016digital, samala2014computer}, have been developed in order to suppress artifacts and obtain improved imaging performance.

%Take the widely used total variation (TV) based IR algorithm as an example, the TV term regularizes pixel intensity variation of the reconstructed image and encourages piecewise constant \cite{mota2015total, sidky2009enhanced}. By doing so, the DBT image noise or image artifacts can be significantly removed. Other types of regularizers can also be utilized\cite{lu2010selective, zheng2016digital, samala2014computer} in DBT image reconstruction. 

Despite of the superior performance, there are several major challenges to use the IR methods in DBT image reconstruction tasks. First, it is hard to manually define an efficient regularization term in DBT to simultaneously reduce artifacts, suppress noise, and preserve image spatial resolution. Second, the reconstructed DBT images may strongly depend on the selections of iteration parameters, for instance, the step size and weighting. Often, these parameters need to be adjusted empirically or according to some common strategies such as L-curve method \cite{doi:10.1137/1034115}, generalized cross-validation (GCV) \cite{doi:10.1080/00401706.1979.10489751}, and Stein's unbiased risk estimate (SURE) \cite{6185677}, which are complicated and require a lot of efforts. Third, the IR methods usually require large computation cost and take long computation time, especially for the 3D DBT reconstruction task.

Over the last few years, the deep learning (DL) technique has attracted a lot of research interests in automatically detecting the microcalcifications \cite{10.1117/12.2217092, doi:10.1118/1.4967345}, lesion classification \cite{MENDEL2019735}, and noise reduction \cite{10.1117/12.2293125, 10.1117/12.2549361} in DBT imaging. In addition, success has also been achieved by implementing the DL technique into the medical image reconstructions, such as the low dose computed tomography (CT) image reconstruction, fast magnetic resonance imaging (MRI) image reconstruction, and so on. Studies\cite{8434327, yang2018admm, cheng2019modelbased} have found that the obtained image quality can be greatly enhanced if unrolling the IR algorithm into the DL network framework, compared with the results obtained from the conventional IR algorithms. In that case, both the regularizer term and iteration parameters are set to be learned from the network. Moriakov \textit{et al.} have tested the DL based primal-dual image reconstruction algorithm\cite{8271999} on the 2D DBT image reconstruction \cite{10.1117/12.2512912}. Results demonstrated that in contrast to the standard iterative reconstruction algorithm, better DBT image quality can be obtained in terms of the quantitative breast density accuracy and the image artifact reduction. However, that network is not compatible with the 3D DBT reconstruction task, which in fact should be more real and meaningful in daily clinical applications.

In order to explore the viability of such DL technique based IR algorithms in 3D DBT imaging applications, we propose a deep iterative reconstruction network called the DIR-DBTnet. Specifically, the well-known alternating direction method of multipliers (ADMM)\cite{boyd2011distributed} is used to solve the DBT image reconstruction problem, and in this study it is unrolled into the DL network via certain strategies. In this particular DIR-DBTnet implemented with the ADMM algorithm, the iteration parameters are assumed as learnable variables, and the regularizer is represented by a network module to extract the complicated 3D prior information through network training. It is worth noting that other popular algorithms, such as the  fast iterative soft thresholding algorithm (FISTA), primal-dual algorithm (PD), split Bregman algorithm, and so on, can all be unrolled into the DL framework to solve the DBT imaging problem.

The remains of this paper are as follows: the basic theory is introduced in Sec. \ref{theory}, and the implementation details of the DIR-DBTnet is given in Sec. \ref{network}. In Sec. \ref{met_and_mat}, the network implementation details, data acquisition procedure and image quality assessment metrics are discussed. The results presented in Sec. \ref{results} demonstrate improved image quality with less in-plane and out-of-plane artifacts. The discussions and conclusion are given in Sec. \ref{dis_con}.

\section{Theory}\label{theory}
The objective function for the DBT image reconstruction problem can be expressed as below:
	\begin{equation}
	\hat{x}=\arg\min_x\frac{1}{2}\left\| Ax-b\right\|_2^2+\beta R(x),
	\label{eq1}
	\end{equation} 
where $x\in\mathbb{R}^{n}$ represents the scanned 3D object; $\hat x\in\mathbb{R}^{n}$ is the estimated image; $b\in\mathbb{R}^{m}$ represents the acquired projection of the object ($m$ is the product of three factors: the number of views $v$, the number of horizontal detector pixels $nd_x$ and vertical detector pixels $nd_y$); $A\in\mathbb{R}^{m\times n}$ denotes the system matrix that models the forward Radon transform; $\left\| \cdot \right\|_2$ denotes the $L_2$ norm; $R(x)$ represents the regularization term; and $\beta$ is a factor that balances the weights of data fidelity term and regularization term. 

For the ADMM algorithm, in general, it divides the complicated objective function into several easy-to-solve subproblems, and then alternatively minimizes each subproblem. As a result, the augmented Lagrangian form of Eq.~(\ref{eq1}) can be rewritten as follows by introducing an auxiliary variable $z = x$:
	\begin{equation}
	L_\rho(x, y, \alpha) = \frac{1}{2}\left\| Ax-b\right\|_2^2+\beta R(z)+
 	\langle \alpha, x-z \rangle +  \frac{\rho}{2}\left\| x-z\right\|_2^2,
	\label{eq2}
	\end{equation} 
where $\alpha$ represents the Lagrangian multiplier, $\rho$ is a penalty parameter. Following the ADMM algorithm, immediately, the corresponding subproblems can be obtained:
	\begin{equation}  
	\label{eq3}
	\left\{  
	             \begin{array} {lr}
	             x^n = \arg\min_x\frac{1}{2}\left\| Ax-b\right\|_2^2+\frac{\rho}{2}\left\| x-z^{n-1}+\lambda^{n-1}\right\|_2^2&  \\  
	             z^n = \arg\min_z \beta R(z)+\frac{\rho}{2}\left\| x^{n}-z+\lambda^{n-1}\right\|_2^2 \\  
	             \lambda^n =\lambda^{n-1}+\eta^n(x^n-z^n)    
	             \end{array}  
	\right.  
	\end{equation} 
where $n\in(1,2,...,N_{iter})$ denotes the ADMM iteration index, $\lambda =  \frac{\alpha}{\rho}$ is the scaled Lagrangian multiplier, and $\eta^n$ is the multiplier update rate. Assuming the subproblems of $x^n$ and $z^n$ are solved by the gradient descent (GD) algorithm, let $k\in(1,2,...,K)$ denote the GD iteration index, thus we have:
\begin{equation}  
\label{eq4}
\left\{  
             \begin{array} {lr}
             X^{n}: x^{n,k}= \mu_1^{n,k}x^{n,k-1}+\mu_2^{n,k}(z^{n-1}-\lambda^{n-1})-\tau^{n,k} A^T(Ax^{n,k-1}-b) &  \\  
             Z^{n}: z^{n,k}= \mu_3^{n,k}z^{n,k-1}+\mu_4^{n,k}(x^{n}+\lambda^{n-1})- \gamma^{n,k} S(z^{n,k-1}) \\  
             \Lambda^n:  \lambda^n= \lambda^{n-1}+\eta^n(x^n-z^n) &    
             \end{array}  
\right.  
\end{equation} 
where $X^n$, $Z^n$ and $\Lambda^n$ represent the reconstruction, denoising and multiplier update modules, respectively; $\mu_1^{n,k}$, $\mu_2^{n,k}$, $\mu_3^{n,k}$, $\mu_4^{n,k}$ and $\gamma^{n,k}$ represent the step size; $S(z)$ represents the gradient of regularization term $R(z)$.

\section{The DIR-DBTnet}\label{network}
In conventional iterative reconstruction algorithms, usually the regularization term and the step size need to be set and adjusted manually. Using the algorithm unrolling strategy proposed by Yang \textit{et al.} \cite{yang2018admm}, the optimization procedure in Eq.~(\ref{eq4}) can be implemented via network in the following way: $S(z^{n,k-1})$ can be realized by a convolutional network (CNN) module that takes $z^{n,k-1}$ as its input; parameters like $\mu_1^{n,k}, \mu_2^{n,k}, \mu_3^{n,k}, \mu_4^{n,k}, \tau^{n,k}, \gamma^{n,k}$ and $\eta^n$ can all be considered as learnable variables in the network. 

	\vspace{0.8cm} 
	\begin{figure}[ht]
	   \begin{center}
	   \includegraphics[width=0.95\textwidth]{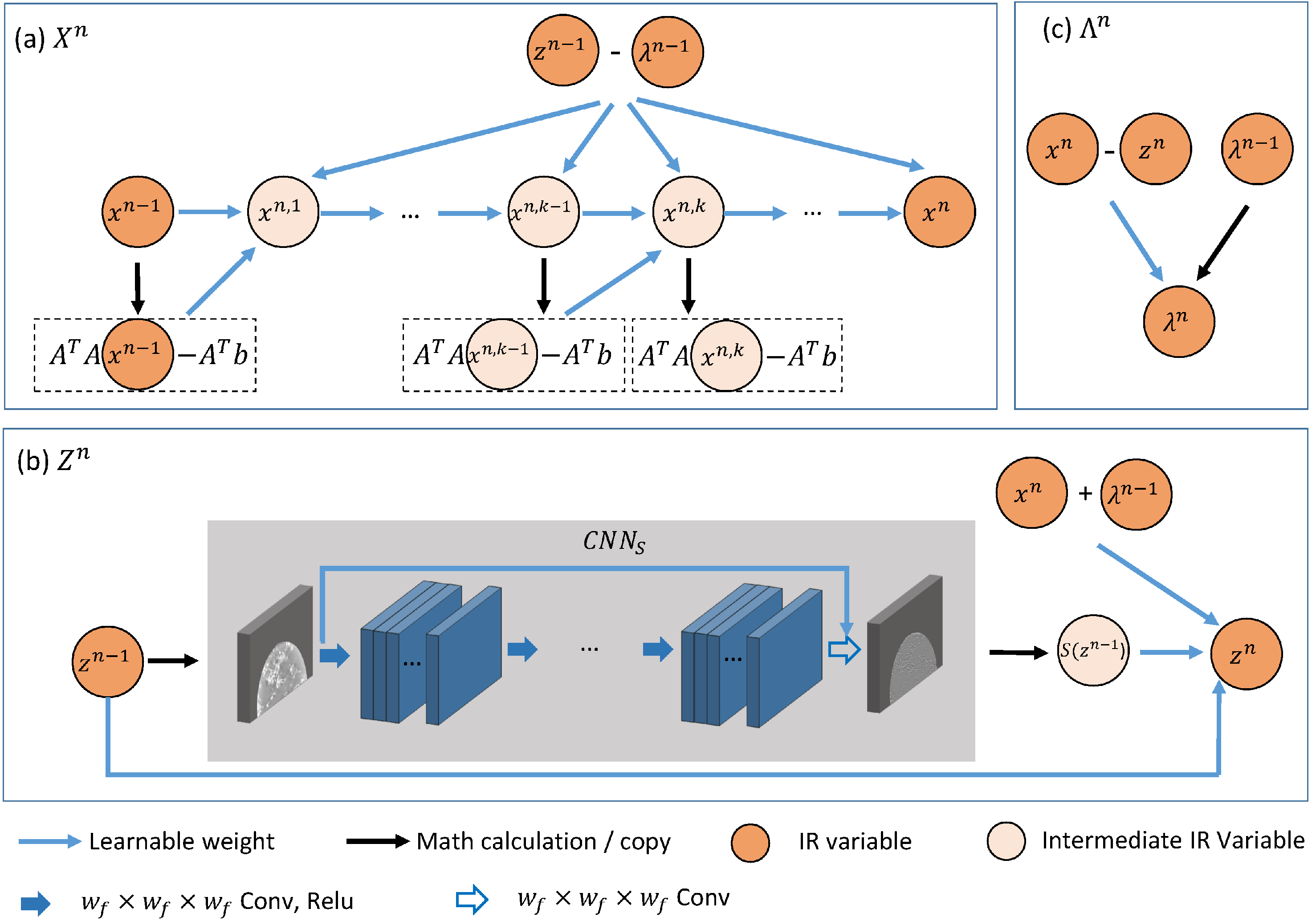}
	   \caption{The network structure and the data flow of the proposed DIR-DBTnet. For the n$_{th}$ iteration loop, details of (a) the $X^n$ module, (b) the $Z^n$ module and (c) the $\lambda^n$ module, are illustrated. For the $Z^n$ module, 3D convolution kernels are used to extract features.
	   \label{fig_Net_scheme} 
	    }  %note label inside caption
	    \end{center}
	\end{figure}
The Fig. \ref{fig_Net_scheme} illustrates the overall scheme of the proposed DIR-DBTnet, including the reconstruction module $X^n$, denoising module $Z^n$ and multiplier update module $\lambda^n$. The iteration parameters to be learned were denoted by blue arrows and the regularizer was represented by CNN$_S$ module. Specifically, due to its training benefits and the outstanding performance in image denoising\cite{chen2017low}, the residual CNN \cite{zhang2017beyond} is utilized to represent the gradient of regularization term $S(z)$. This CNN module is denoted as CNN$_S$. There are $l$ hidden layers in it, and 3D convolution kernel with size of $w_f\times w_f\times w_f$ is used for each layer to extract both in-plane and out-of-plane prior information. The shortcut connection between input and output data enforces the CNN$_S$ module to learn the residual. Additionally, details of the algorithm implementation are listed in Table \ref{table_I}.

	\begin{table}[h]
	\caption{Algorithm implementation of DIR-DBTnet}
	\label{table_I}
	\centering
	\begin{tabular}{l}
	\hline
%	Algorithm implementation of DIR-DBTnet \\
%	\hline
	Initialization: $x =0$, $z = 0$, $\lambda=  0$ \\
	For $n = 1:N_{iter}$, do \\
	\ \ \ \ \ For $k = 1:K_x$, do \\
	\ \ \ \ \ \ \ \ \ \ $x \leftarrow \mu_1^{n,k}x+\mu_2^{n,k}(z-\lambda)-\tau^{n,k} A^T(Ax-b)$ \\
	\ \ \ \ \ end \\
	\ \ \ \ \ For $k = 1:K_z$, do \\
	\ \ \ \ \ \ \ \ \ \  $S(z) \leftarrow$ CNN$_S(z)$\\
	\ \ \ \ \ \ \ \ \ \ $z \leftarrow \mu_3^{n,k}z+\mu_4^{n,k}(x+\lambda)-\gamma^{n,k} S(z) $ \\
	\ \ \ \ \ end \\
	\ \ \ \ \ $\lambda \leftarrow \lambda+\eta^n(x-z) $ \\
	end\\
	(Note: The $\mu_1^{n,k}, \mu_2^{n,k}, \mu_3^{n,k}, \mu_4^{n,k}, \tau^{n,k}, \gamma^{n,k}, \eta^n$, weights,\\ and biases in CNN$_S$  are all initialized and updated\\ automatically during the network training.) \\
	\hline
	\vspace{0.0cm} 
	\end{tabular}
	\end{table}

During the iterative reconstruction procedure, apparently, the matrix $A$ and its adjoint $A^T$ that correspond to the forward projection (FP) and backprojection (BP) need to be calculated repeatedly. Since the DBT imaging system uses the cone-beam geometry to acquire the 3D projections, the $A$ and its adjoint $A^T$ would consume large amount of computation resource if stored on RAM. To address this problem, we developed two user-defined GPU accelerated TensorFlow operators: $\mathbb{FP}(\cdot)$ and $\mathbb{BP}(\cdot)$. In particular, the two operators accept a group of system geometry parameters\cite{gao2012fast} and finally output the forward projection and backprojection results. All the calculations of $\mathbb{FP}(\cdot)$ and $\mathbb{BP}(\cdot)$ operators are implemented in parallel by CUDA.

\section{Methods and Materials} \label{met_and_mat}
\subsection{Training and testing dataset} \label{dataset generation}
In this study, the large amount of data required by network training is numerically simulated, rather than being collected from clinical centers. This is because the images generated by the commercially available DBT systems are not ideal and still contain undesired artifacts, along both in-plane and out-of-plane directions. To demonstrate the advancement of the DIR-DBTnet in improving the DBT imaging performance, it would be highly desired to train the network with DBT images that have no artifacts. Therefore, numerically simulated DBT sinogram-label pairs are generated and used.

	\vspace{0.8cm} 
	\begin{figure}[t]
	   \begin{center}
	   \includegraphics[width=0.8\textwidth]{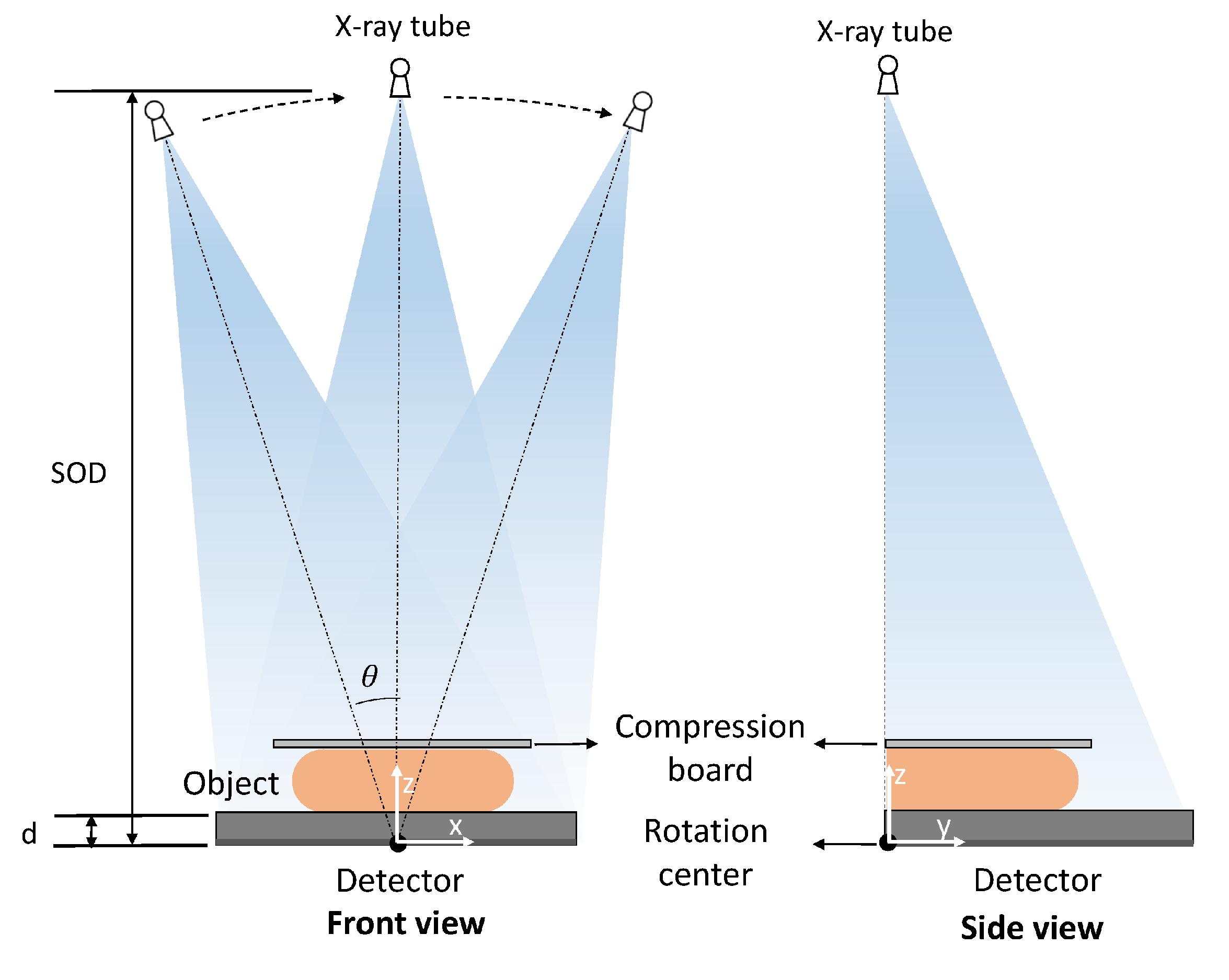}
	   \caption{Illustration of the DBT imaging system.
	   \label{fig_acquisition} 
	    }  %note label inside caption
	    \end{center}
	\end{figure}
The simulated DBT imaging system used in this study has the same geometry as a commercial one (Model: Mammo 890i, United Imaging, Shanghai, China), see the illustration in Fig.~\ref{fig_acquisition}. The distance from the X-ray source to the detector is set to 700 mm, and the bottom of the phantom is 20 mm away from the detector. In total, 21 projections are obtained from an angular tube rotation range of $-20^{\circ}$ to $20^{\circ}$, with an angular interval of $2^{\circ}$. The CMOS based indirect-conversion type X-ray detector has an active matrix size of $2945\times2304$, and the native detector element dimension is $99\;\mu m$. When acquiring the projections, the detector works in the $2\times2$ binning mode. No anti-scatter grid is added.

\begin{figure}[htb]
	   \begin{center}
	   \includegraphics[width=1\textwidth]{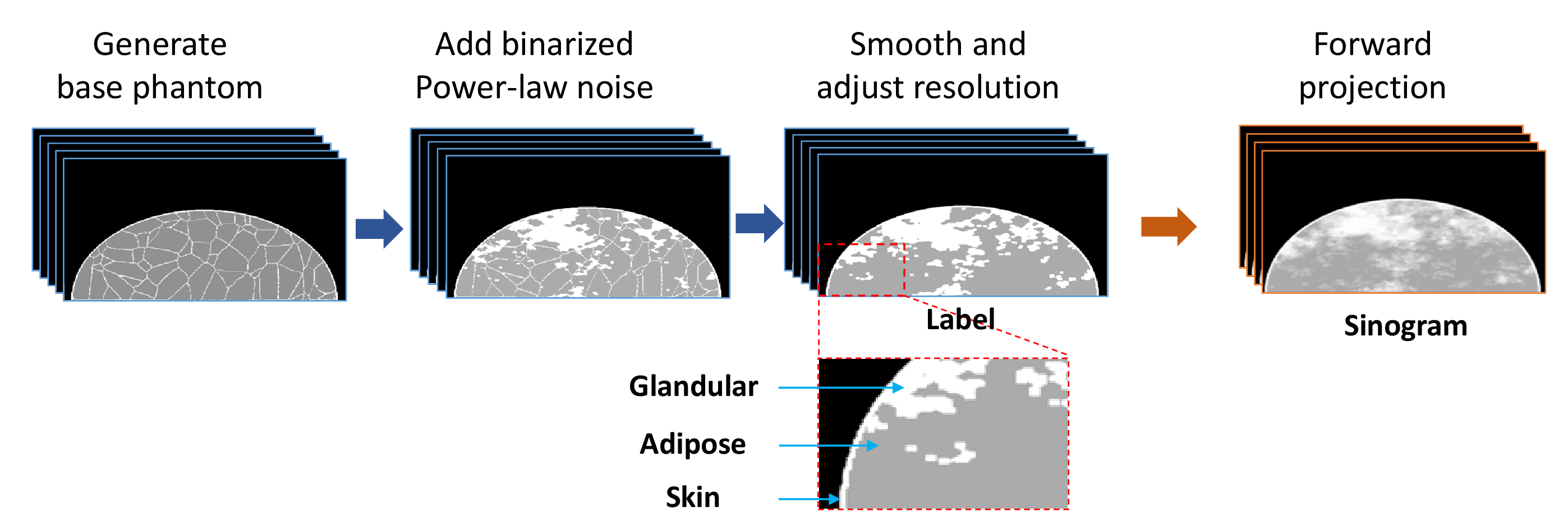}
	   \caption{Workflow to generate the numerical breast phantom. First, the base breast phantom is created, and a shell of skin and adipose compartments are included; Second, fine fibroglandular structures are simulated by adding the binarized power-law noise into certain regions; Third, the image is smoothed to avoid rough edges and then adjusted to the desired phantom resolution; Finally, forward projections of the digital phantom are simulated.
	   \label{fig_training_data} 
	    }  %note label inside caption
	    \end{center}
	\end{figure}

Moreover, the simulated breast phantom is made of skin, adipose, and glandular tissues. The attenuation coefficients of adipose and glandular were set to 0.0456 mm$^{-1}$ and 0.0802 mm$^{-1}$, respectively\cite{Johns_1987}. The attenuation coefficient of skin is set equal to 0.0802 mm$^{-1}$ as well. Details of these numerical breast simulations\cite{doi:10.1118/1.4718576} are illustrated in Fig.~\ref{fig_training_data}. Herein, the simulated digital breast phantom has a matrix of $512\times1024\times30$, with the voxel dimension of 0.15 mm $\times$0.15 mm$\times$1.75 mm. The simulated detector has $512\times1024$ pixels with each pixel dimension of 0.198 mm$\times$0.198 mm. To increase the data diversity, the shape and size of the simulated breast phantom are set to vary randomly: the volumetric glandular fraction changes from 13.7\% to 25.6\%. In total, 1000 digital phantoms are synthesized, and their forward projections are generated correspondingly. Additionally, 100 pairs of projection-label data are generated as testing dataset following the same procedure, except for that small round high-contrast objects with various dimensions are added to mimic the microcalcifications.

	\vspace{0.5cm}

\subsection{Network training}
To train the DIR-DBTnet, several hyper-parameters need to be determined in prior. The first one is the number of ADMM iterations $N_{iter}$. According to the reference \cite{he2018optimizing}, we set $N_{iter}=10$ with the purpose to balance both the computation cost and the reconstruction quality. The second hyper-parameter is the number of gradient descent iterations $K_x$ and $K_z$ when solving $x^n$ and $z^n$. Empirically, we set $K_x=5$ and $K_z=1$. In addition, the CNN$_S$ module consists of 2 hidden layers, which use 3D convolution kernels with size of $ w_f=3$, and filter number of $N_f=16$, and the final output layer adopts a single convolution filter with the same kernel size. 

The network was trained to minimize the standard mean square error (MSE) loss between the 3D network output and the label images in Tensorflow using Adam algorithm. The starting learning rate was set to $1\times10^{-5}$ with exponentially decaying rate of 0.98 after every 1000 steps. The network was trained for 30 epochs with batch size of 1 on single INVIDIA TITAN RTX GPU card. The entire training process took about 192 hours.

\subsection{Experiment}
The BR3D breast phantom (Model: 020, CIRS, VA, USA) was scanned using the United Imaging DBT imaging system. The phantom contains fibers, specks, and masses of various dimensions to simulate microcalcifications, fibrous structures and tumors. It also consists of complex swirls of simulated adipose and glandular tissues, making it more challenging for target detection. The 2.0 cm thick BR3D phantom was scanned with tube voltage of 26 kVp and tube current of 100 mA.

\subsection{Quantitative evaluation metrics}
To quantitatively evaluate the DBT imaging performance of the proposed DIR-DBTnet, both the artifact spread function (ASF) and the signal difference to noise ratio (SDNR) are used. In particular, the ASF quantifies the out-of-plane artifact:
	\begin{equation}
	\rm{ASF}(z) = \frac{\overline\mu_{Artifact}(z)-\overline\mu_{BG}(z)}{\overline\mu_{Feature}(z_0)-\overline\mu_{BG}(z_0)},
	\label{eq5}
	\end{equation} 
where $z$ denotes the slice depth along $z$ axis, $z_0$ denotes the depth of the feature of interest, $\overline\mu_{\rm{Feature}}(z_0)$ denotes the average feature intensity value, $\overline\mu_{\rm{Artifact}}$ represents the average intensity value of the artifact region, and $\overline\mu_{\rm{BG}}$ denotes the adjacent background region. The full width half maximum (FWHM) of the obtained ASF curve is also measured. 

Moreover, the SDNR quantifies the in-plane detectability of features, and is defined as follows: 
	\begin{equation}
	\rm{SDNR} = \frac{\overline\mu_{Feature}-\overline\mu_{BG}}{\sigma_{BG}},
	\label{eq6}
	\end{equation} 
where $\sigma_{BG}$ denotes the standard deviation of the pixel values in the background region.

\section{Results}\label{results}
\subsection{Numerical simulation results}
\begin{figure}[ht]
	   \begin{center}
	   \includegraphics[width=1\textwidth]{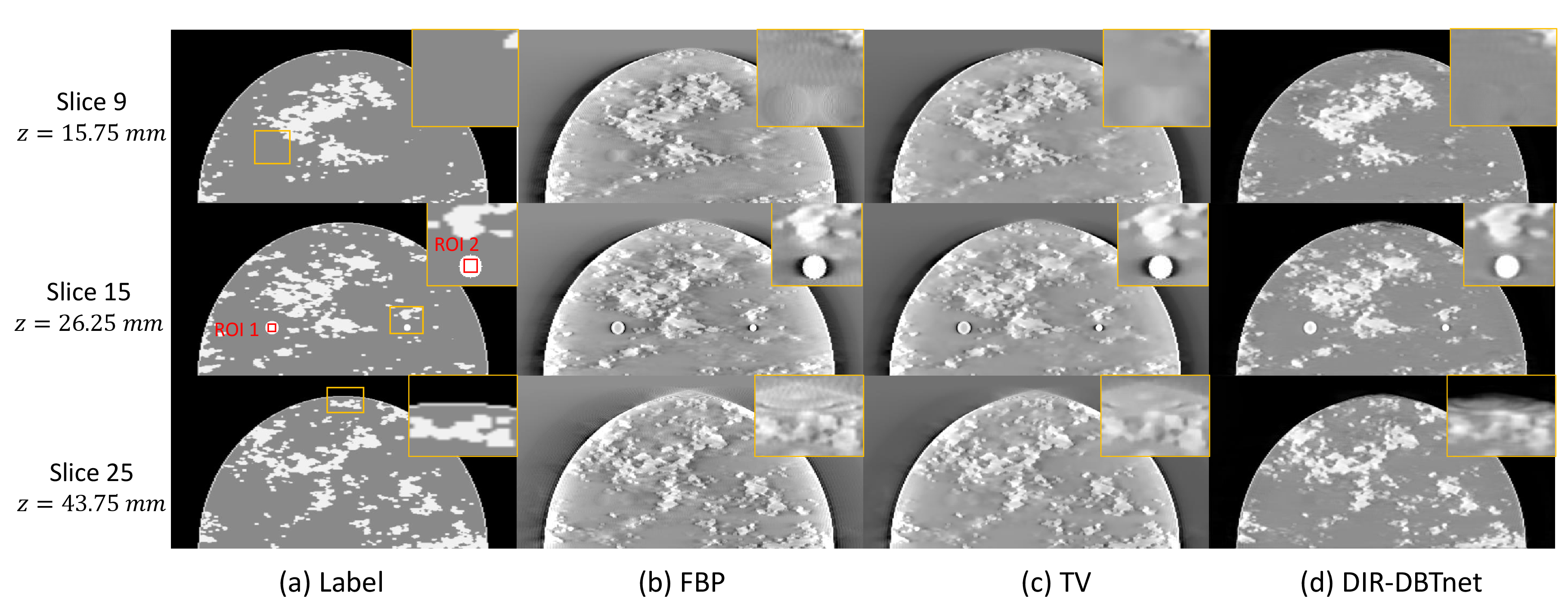}
	   \caption{Reconstruction results of the testing dataset using different methods: (a) Label, (b) FBP, (c) TV, (d) DIR-DBTnet. The rows from top to bottom represent different slice positions along $z$ direction. Note that images are displayed with different windows to generate similar gray levels for the adipose tissue.
	   \label{fig_results_simu} 
	    }  %note label inside caption
	    \end{center}
	\end{figure}

Fig. \ref{fig_results_simu} shows the reconstruction results of the simulated numerical breast phantom. Clearly, the shadow artifacts around the glandular structures and the phantom peripheries in the FBP and TV reconstructed images are more dramatic than the images reconstructed by the DIR-DBTnet. For the high contrast calcification in the selected region-of-interest (ROI), the FBP reconstruction algorithm leads to serious horizontal dark shadows, see the magnified ROI 2 on slice 15. Such shadows are slightly mitigated by the TV method, while our proposed DIR-DBTnet method removes almost all the artifacts, and generates similar DBT images as the label images. From the magnified ROI images on slice 25, it can be observed that the blurred phantom boundaries reconstructed by the FBP and TV algorithms get much clearer in the DIR-DBTnet method. 
The magnified ROI images on slice 9 show the out-of-plane artifacts caused by the ROI 1 on slice 15, and the proposed DIR-DBTnet presents the fewest artifacts as well. Overall, results show that the DIR-DBTnet outperforms the FBP and TV methods in three aspects: fewer in-plane shadow artifacts, more accurate phantom boundaries, and reduced out-of-plane artifacts.
	\vspace{0.1cm}
	
	\begin{figure}[htb]
	\begin{minipage}[b]{0.5\linewidth}
	\centerline{\includegraphics[width=8cm]{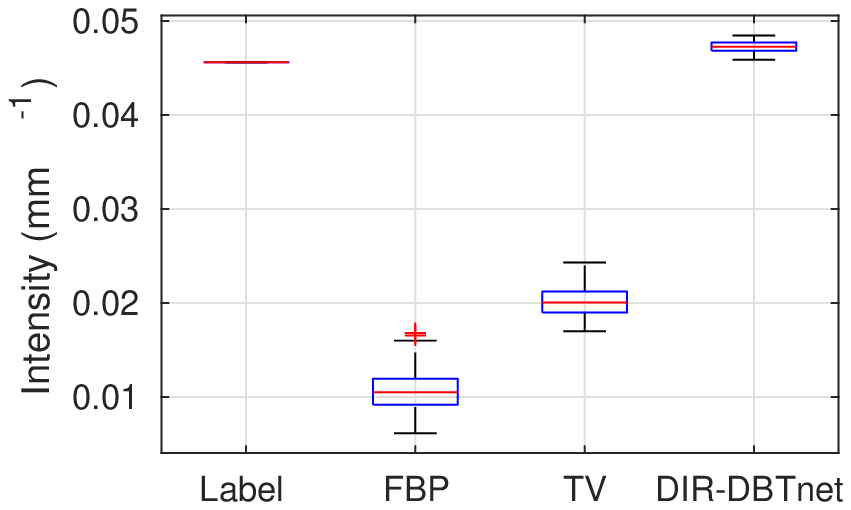}}
	 \centerline{(a)}
	\end{minipage}
	\begin{minipage}[b]{0.49\linewidth}
	\centerline{\includegraphics[width=8cm]{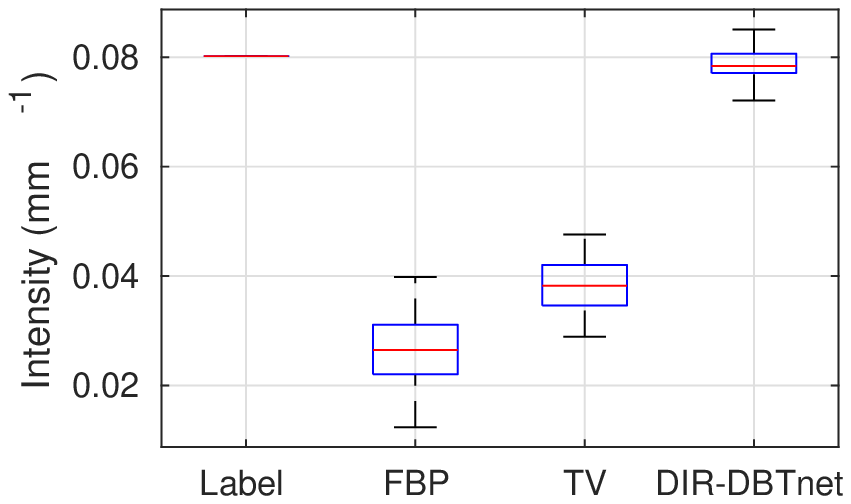}}
	 \centerline{(b)}
	\end{minipage}
	 \caption{The box plot results of the measured breast tissue density: (a) the adipose and (b) the glandular tissue. Measurements were performed on two ROIs corresponding to the adipose and the glandular tissue respectively, each selected ROI contains 25$\times$25 pixels. }
	\label{fig_w_b}
	\end{figure}

The breast density (attenuation coefficient) comparison results of the adipose and the glandular tissues are plotted on Fig. \ref{fig_w_b} for different DBT image reconstruction algorithms. Clearly, the proposed DIR-DBTnet is able to significantly improve the image accuracy, while the breast densities obtained using the FBP and TV methods are strong deviated from the ground truth. With such improved signal accuracy, the DIR-DBTnet method can help to enhance the quantitative DBT imaging performance.

	\begin{figure}[htb]
	\begin{minipage}[b]{0.5\linewidth}
	\centerline{\includegraphics[width=8.5cm]{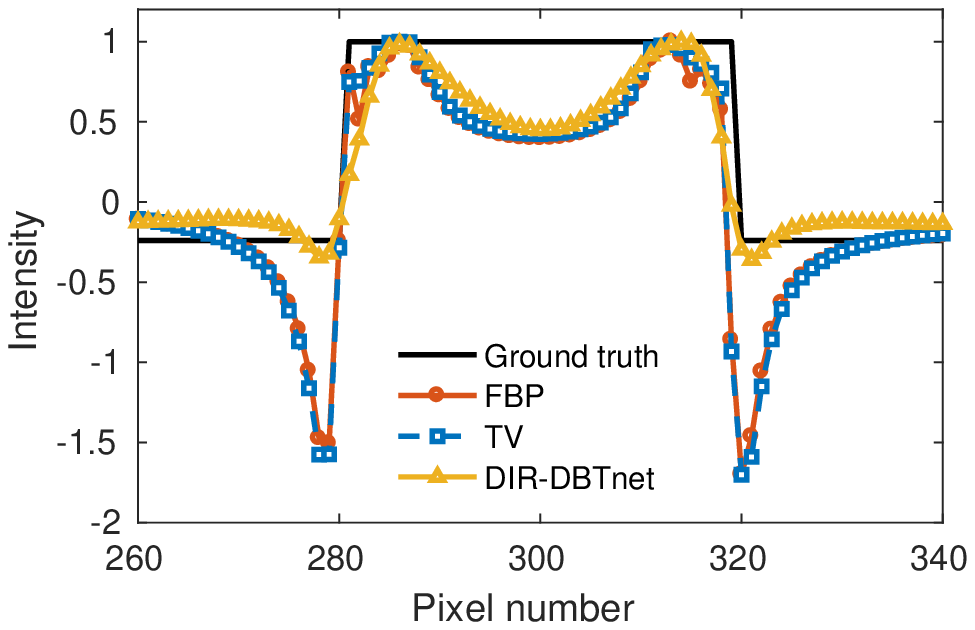}}
	 \centerline{(a)}
	\end{minipage}
	\begin{minipage}[b]{0.48\linewidth}
	\centerline{\includegraphics[width=8.5cm]{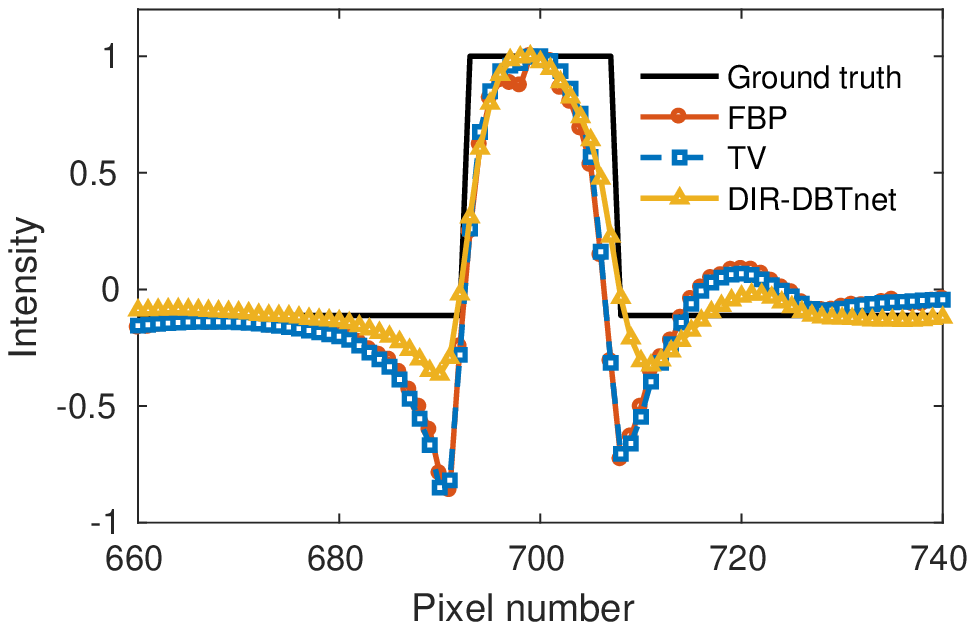}}
	 \centerline{(b)}
	\end{minipage}
	 \caption{Results of the normalized line-profiles of the small calcification: (a) on the ROI 1, (b) on the ROI 2, highlighted in Fig. \ref{fig_results_simu}.}
	\label{fig_profile_simu}
	\end{figure}

The quantitative assessment results of the in-plane image quality are shown in Fig.~\ref{fig_profile_simu}. They represent the normalized line-profiles \cite{wu2004comparison} of small calcifications in ROI 1 and ROI 2 highlighted in Fig.~\ref{fig_results_simu}. As seen, the DIR-DBTnet could suppress the shadow artifacts of the edges efficiently, and thus generates the most consistent profiles which are close to the ground truth. Whereas, both the FBP and TV algorithms exhibit strong drops at the feature edges, indicating severe shadow artifacts. Moreover, the small rise on the right side of the curve (around the 720-th pixel) in Fig. \ref{fig_profile_simu}(b) corresponds to the artifacts caused by features in neighboring slices, but they are mitigated by the DIR-DBTnet algorithm.

	\begin{figure}[htb]
	\begin{minipage}[b]{0.5\linewidth}
	\centerline{\includegraphics[width=8.5cm]{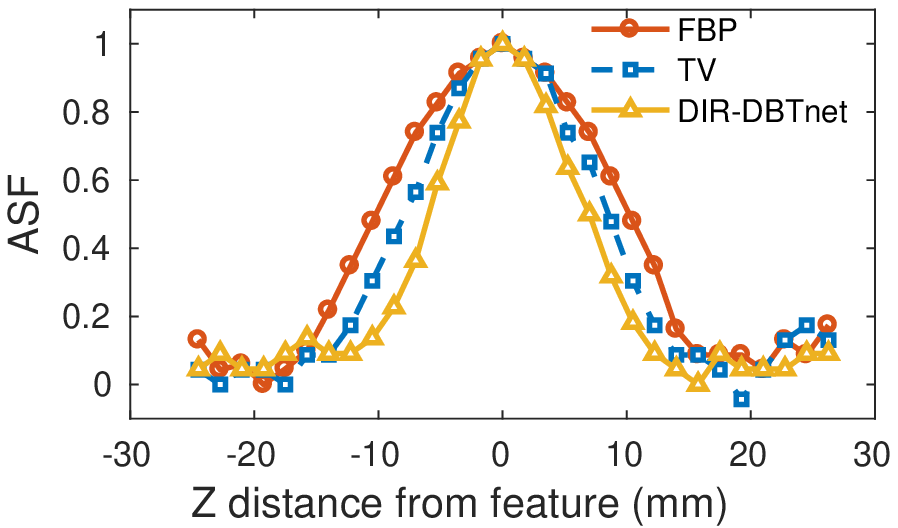}}
	 \centerline{(a)}
	\end{minipage}
	\begin{minipage}[b]{0.49\linewidth}
	\centerline{\includegraphics[width=8.5cm]{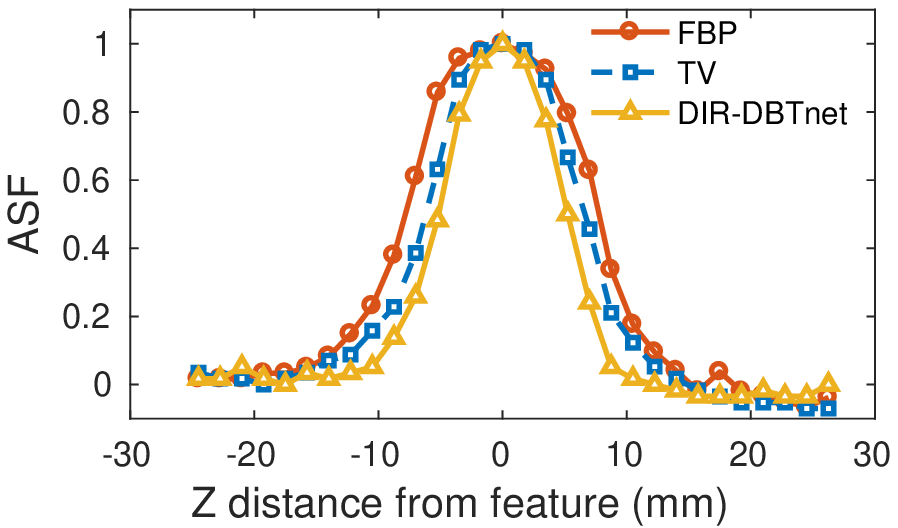}}
	 \centerline{(b)}
	\end{minipage}
	 \caption{Measured ASF curves for different reconstruction methods: (a) ROI 1 and (b) ROI 2, as marked by red rectangles in Fig. \ref{fig_results_simu}. A neighboring area beneath the feature or artifact location was chosen as background region for the calculation of ASF. }
	\label{fig_ASF}
	\end{figure}

To quantify the out-of-plane artifacts, the ASFs of ROI1 and ROI2 are plotted. Results are shown in Fig.~\ref{fig_ASF}. The FBP algorithm produces a slowly decreasing curve as the $z$ distance increases from the interested feature, corresponding to severe out-of-plane artifacts. The TV method shows a slightly better result, and the ASF curve gets a little narrower. The DIR-DBTnet generates the narrowest ASF curve, namely, it has the best capability in suppressing the out-of-plane artifacts. Quantitatively, the measured FWHMs of the FBP, TV, and DIR-DBTnet methods are 19.5 mm, 16.0 mm and 12.9 mm for Fig.~\ref{fig_ASF}(a), and are 15.2 mm, 12.7 mm and 10.2 mm for Fig.~\ref{fig_ASF}(b). On average, the DIR-DBTnet obtains approximately 33.4\% and 19.7\% improvements compared to FBP and TV methods, correspondingly.

\subsection{Experimental results}
\begin{figure}[htb]
	   \begin{center}
	   \includegraphics[width=15cm]{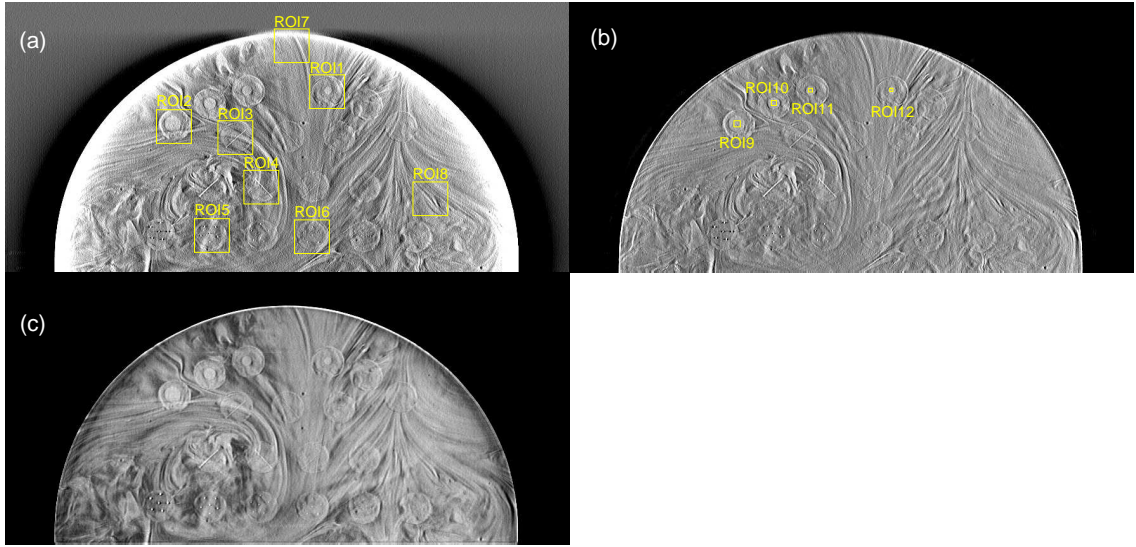}
	   \caption{Reconstruction results of the same slice of BR3D phantom using different methods: (a) FBP method, (b) TV method, and (c) DIR-DBTnet.
	   \label{fig_results_BR3D} 
	    }  %note label inside caption
	    \end{center}
	\end{figure}

\begin{figure}[htb]
	   \begin{center}
	   \includegraphics[width=15.5cm]{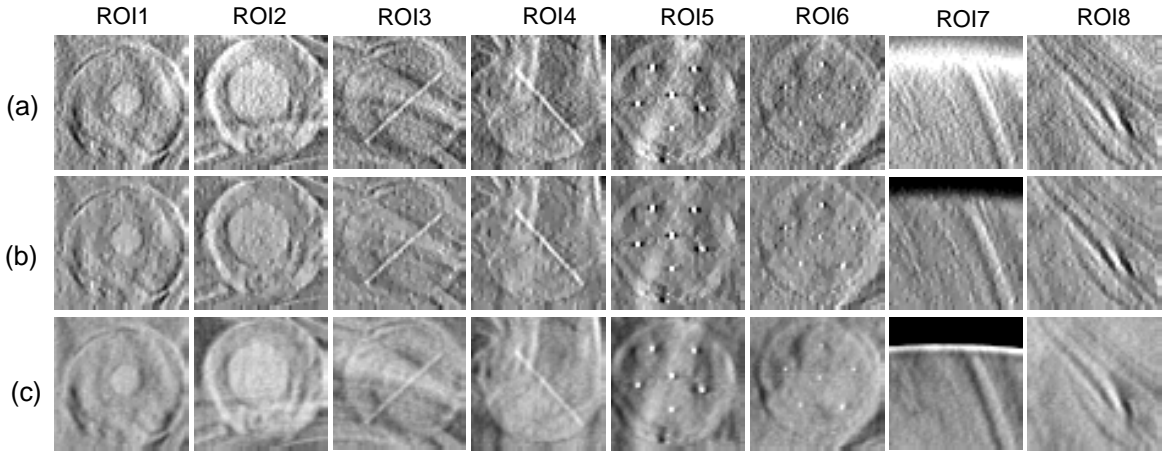}
	   \caption{Magnified ROI regions in Fig. \ref{fig_results_BR3D} reconstructed by (a) FBP method, (b) TV method, and (c) DIR-DBTnet.
	   \label{fig_results_BR3D_ROI} 
	    }  %note label inside caption
	    \end{center}
	\end{figure}

The reconstructed experimental results (one certain slice) of the breast phantom are shown in Fig. \ref{fig_results_BR3D}. The images have a size of $512\times1024$. Both the TV method and DIR-DBTnet are able to reduce the shadow artifacts around the phantom periphery. Moreover, the generated images from the proposed DIR-DBTnet method are less noisy.  Results in Fig. \ref{fig_results_BR3D_ROI} show the magnified ROI images for the selected ROIs, including masses (ROI1 and ROI2), fibrils (ROI3 and ROI4), specks (ROI5 and ROI6), boundaries (ROI7) and swirl patterns (ROI8). Features in these selected regions are well preserved after reconstructed by the DIR-DBTnet, while noises are suppressed efficiently as well. 

	\vspace{0.2cm}

\begin{figure}[htb]
   \begin{center}
   \includegraphics[width=16cm]{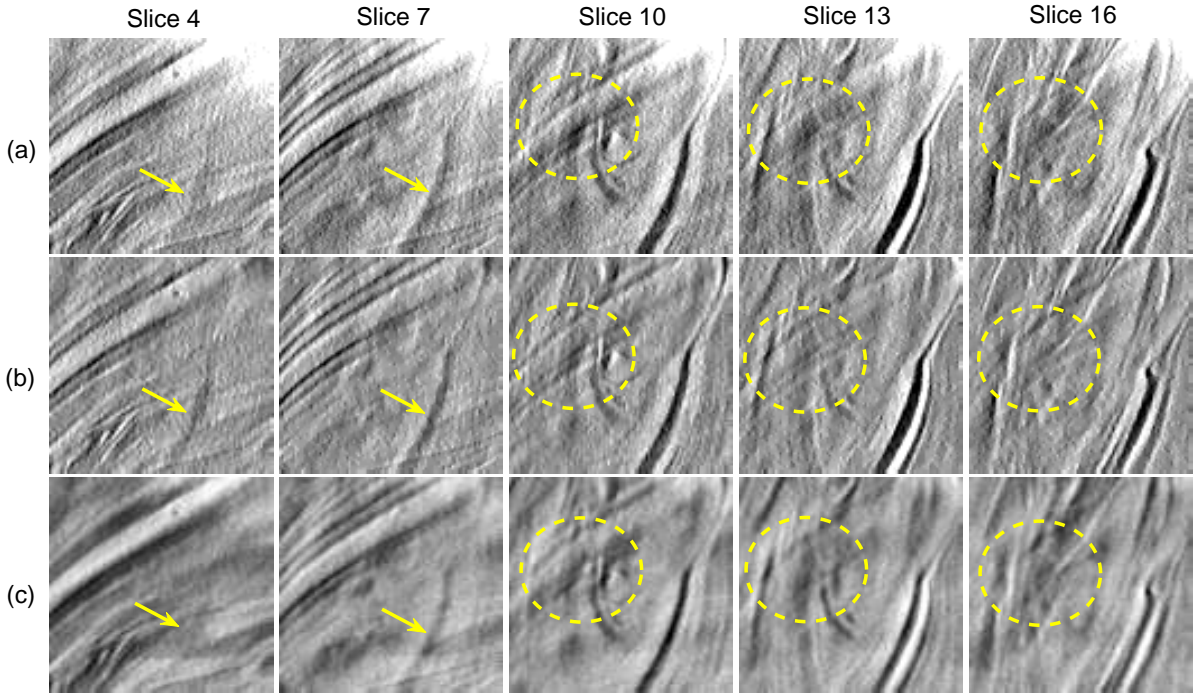}
   \caption{The DBT images at different $z$ depths reconstructed by: (a) FBP, (b) TV, (c) DIR-DBTnet. The strips pointed by arrows are the artifacts spread from other slices. The ellipses highlight the streak artifacts caused by the high-contrast fibrous textures in other slices.
   \label{fig_outplane_br3d} 
    }  %note label inside caption
    \end{center}
\end{figure}

Fig. \ref{fig_outplane_br3d} illustrates the out-of-plane artifacts by showing the magnified images with the same $xy$ position but different $z$ slices. The dark strip denoted by arrows and the streak textures denoted by circles are all artifacts caused by the high-contrast structures in latter or former slices. It can be observed that these artifacts were better suppressed in (c) by the DIR-DBTnet method.

\begin{figure}[htb]
   \begin{center}
   \includegraphics[width=12cm]{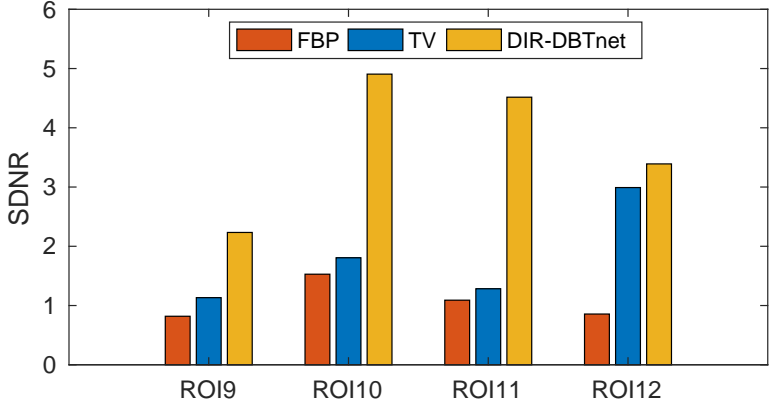}
   \caption{Comparison of SDNRs of ROI9 to ROI12 marked in Fig. \ref{fig_results_BR3D}(b) using different reconstruction methods. A neighboring area beneath each ROI was chosen as background region for the calculation of SDNR.
   \label{fig_reSDNR_simu} 
    }  %note label inside caption
    \end{center}
\end{figure}

The results in Fig.~\ref{fig_reSDNR_simu} show the comparison of the SDNR measurements. Due to the reduced image noise, it can be seen that the DIR-DBTnet generates the highest SDNR values. On average, such improvement of SDNR is 250.1\% and 133.5\% compared to FBP and TV methods, respectively.

In addition, the reconstruction time spent by the DIR-DBTnet is also much shorter than the TV method. In the present work, reconstructing a $512\times1024\times30$ DBT image volume from the $512\times1024\times41$ projection data takes about 9.3 s. As a contrary, the TV method needs 115.0 s for 50 iterations under the same computation condition.

%Figure. \ref{fig_profile_br3d} shows the normalized line-profiles of the mass within ROI2 and the central speck within ROI5 in Fig. \ref{fig_results_BR3D}. The profiles obtained from the DIR-DBTnet show less fluctuations than the other two methods, indicating less noises, see Fig. \ref{fig_profile_br3d}(a). Moreover, the sudden drops around the feature edges that reconstructed from the FBP and TV algorithms are also mitigated when reconstructed by the developed DIR-DBTnet, see Fig. \ref{fig_profile_br3d}(b). 

%	\begin{figure}[htb]
%	\begin{minipage}[b]{0.5\linewidth}
%	\centerline{\includegraphics[width=9cm]{Profile_mass.eps}}
%	 \centerline{(a)}
%	\end{minipage}
%	\begin{minipage}[b]{0.49\linewidth}
%	\centerline{\includegraphics[width=9cm]{Profile_speck.eps}}
%	 \centerline{(b)}
%	\end{minipage}
%	\captionv{16}{}{Normalized line-profiles of the mass within ROI2 (a) and of the central speck within ROI5 (b) in Fig. \ref{fig_results_BR3D}.}
%	\label{fig_profile_br3d}
%	\end{figure}

\section{Discussions and Conclusion} \label{dis_con}
In this work, we have demonstrated the viability of using deep learning technique (DIR-DBTnet) to reconstruct the 3D DBT images, with reduced in-plane and out-of-plane artifacts, and improved breast density accuracy compared to the FBP and TV methods. By unrolling the optimization procedure of the iterative reconstruction algorithm into network structures, the tedious parameter tuning process can be avoided, and more efficient regularizer which extracts 3D prior information of image can be learned. We also show that the network trained by the numerically simulated DBT data could be used directly onto the experimentally acquired DBT data and achieved high quality reconstruction with less image artifacts. 

For this 3D DBT image reconstruction task, we proposed to utilize the 3D convolution kernels in the DIR-DBTnet. One benefit of doing so is that both the in-plane and out-of-plane prior information can be learned simultaneously by the deep neural network. Thus, artifacts along the two directions could be well mitigated at the same time. We also tried the 2D convolution kernels merely within the $xy$ plane (slice-wise regularization\cite{doi:10.1002/mp.12864, mota2015total}). The network performance was degraded due to the increased out-of-plane artifacts and the inconsistent intensity distributions along the $z$ direction. Therefore, the 3D convolution kernels should be recommended for DIR-DBTnet to improve the network reconstructed image quality. 

Unlike the other deep learning based medical image reconstruction tasks, for instance, CT, MRI, and so on, it is very difficult to collect high quality paired DBT data from clinical modality and use them to train the DIR-DBTnet. This is mainly because there are no artifact-free DBT images that can be used as network labels, even though the projections can be collected easily. As a consequence, we proposed to generate training data for the network by numerical simulations. So far, we thought this is a very efficient way to prepare training data for DIR-DBTnet. Fortunately, experiments have verified the effectiveness of this data preparation strategy. However, such strategy may become less accurate for real clinical applications, since the simulated data may lack diversity in both tissue structures and densities. To partially compensate this potential drawback, the MRI breast images or the dedicated breast CT images would become the alternative options that can be used as the artifact-free labels. Inevitably, certain image processing should be considered in advance to convert them into the DBT type images. For example, the true 3D breast volumetric images need to be compressed properly to mimic the breast compression effect. In addition, the 3D MRI breast images need to be segmented, and the values need to be converted into X-ray attenuation coefficients. All these work are interesting ones to us and will be investigated in future.

The present work, as a preliminary attempt to reconstruct 3D DBT images using deep learning technique, still has room for improvement in future. First, the size of the reconstructed DBT images should be increased. Currently, due to our restricted GPU memory size (24 GB),  the network generated DBT images have smaller size than those of clinical use. Second, the number of training data could be increased. The present network is trained with only 1000 training samples, and augmentation of the sample volume may improve the DBT reconstruction performance of the DIR-DBTnet. Third, the architecture of the CNN$_S$ module could be optimized. Now we use a simple residual network with 2 hidden layers for the CNN$_S$ module, and increment of the number of hidden layers or optimization of network structure could enhance the DIR-DBTnet performance. Finally, the loss function of the network can be optimized. Although artifacts have been well suppressed using the very standard MSE type loss, the output image of DIR-DBTnet was a bit smooth compared with the images obtained from the FBP and TV methods. As demonstrated in literatures \cite{doi:10.1002/mp.13713, ge2020adaptive}, more advanced loss functions, such as the VGG perceptual loss and the WGAN loss, may encourage edge preservation and thus yield more natural images.

%In summary, based upon our findings that deep learning technique helps to reduce both in-plane and out-of-plane artifacts for 3D DBT reconstruction, some further investigations on more generalized network reconstruction models, dual-energy DBT imaging etc. could be of research interest.

\section{Acknowledgment}
This project was supported by the Guangdong Basic and Applied Basic Research Foundation (Grant No. 2019A1515011262), National Natural Science Foundation of China (Grant No. 11804356), and Chinese Academy of Sciences Key Laboratory of Health Informatics (Grant No.~2011DP173015).

\bibliography{refs}
\end{document}